\documentclass[12pt,a4paper]{iopart}
\usepackage[utf8]{inputenc} 
\usepackage{iopams}
\usepackage{textcomp}
\usepackage{graphicx}
\usepackage[dvipsnames]{xcolor}
\usepackage{soul} 
\usepackage[numbers,comma,square,sort&compress,merge]{natbib}
\usepackage{braket}

\usepackage[
,textwidth=15cm
,textheight=20cm
,verbose
,dvips
]{geometry}

\newcommand{\text}[1]{%
{\mbox{\scriptsize #1}}%
}%

\newcommand{\mtext}[1]{%
{\mbox{\normalsize #1}}%
}%

\usepackage[pdftex,unicode=true,bookmarks=false,breaklinks=false,pdfborder={0 0 1},colorlinks=true]{hyperref}
\hypersetup{pdftitle={Axial GaAs/Ga(As,Bi) Nanowire Heterostructures},linkcolor=blue,citecolor=blue,urlcolor=blue}

\begin{document}

\title[Axial GaAs/Ga(As,Bi) Nanowire Heterostructures]{Axial GaAs/Ga(As,Bi) Nanowire Heterostructures}

\author{Miriam Oliva}
\address{Paul-Drude-Institut für Festkörperelektronik, Leibniz-Institut im Forschungsverbund Berlin e.\,V., Hausvogteiplatz 5--7, 10117 Berlin, Germany}
\author{Guanhui Gao}
\address{Paul-Drude-Institut für Festkörperelektronik, Leibniz-Institut im Forschungsverbund Berlin e.\,V., Hausvogteiplatz 5--7, 10117 Berlin, Germany}
\address{Present address: Material Science and NanoEngineering department, Rice University, 6100 Main Street MS 364, Houston, TX 77005, USA}
\author{Esperanza Luna, Lutz Geelhaar}
\address{Paul-Drude-Institut für Festkörperelektronik, Leibniz-Institut im Forschungsverbund Berlin e.V., Hausvogteiplatz 5–7, 10117 Berlin, Germany}
\author{Ryan B. Lewis}
\address{Paul-Drude-Institut für Festkörperelektronik, Leibniz-Institut im Forschungsverbund Berlin e.V., Hausvogteiplatz 5–7, 10117 Berlin, Germany}
\address{Department of Engineering Physics, McMaster University, L8S 4L7 Hamilton, Canada}
\ead{oliva@pdi-berlin.de}

\begin{abstract}
Bi-containing III-V semiconductors constitute an exciting class of metastable compounds with wide-ranging potential optoelectronic and electronic applications. However, the growth of III-V-Bi alloys requires group-III-rich growth conditions, which pose severe challenges for planar growth. In this work, we exploit the naturally-Ga-rich environment present inside the metallic droplet of a self-catalyzed GaAs nanowire to synthesize metastable GaAs/GaAs$_{1-\text{x}}$Bi$_{\text{x}}$ axial nanowire heterostructures with high Bi contents. The axial GaAs$_{1-\text{x}}$Bi$_{\text{x}}$ segments are realized with molecular beam epitaxy by first enriching only the vapor-liquid-solid (VLS) Ga droplets with Bi, followed by exposing the resulting Ga-Bi droplets to As$_2$ at temperatures ranging from 270 to 380$\,^{\circ}$C to precipitate GaAs$_{1-\text{x}}$Bi$_{\text{x}}$ only under the nanowire droplets. Microstructural and elemental characterization reveals the presence of single crystal zincblende GaAs$_{1-\text{x}}$Bi$_{\text{x}}$ axial nanowire segments with Bi contents up to (10$\pm$2)$\%$. This work illustrates how the unique local growth environment present during the VLS nanowire growth can be exploited to synthesize heterostructures with metastable compounds.
\end{abstract}

\maketitle

\section{Introduction}
Alloying III-V semiconductors with Bi—the group V element which is the heaviest of all non-radioactive elements—has been a hot topic in recent years. This is largely due to the huge reduction of the III-V host bandgap energy per unit Bi incorporation compared to conventional/lighter alloying elements, making III-V-Bi alloys promising for extending the wavelength range of optoelectronic devices for a given III-V material system \cite{francoeur2003band,Marko2012,zhong2012effects,rajpalke2014high,sandall2014demonstration,masnadi2014bandgap}. Owing to the large Bi mass, Bi-containing materials are also of interest for spintronics \cite{fluegel2006giant,mazzucato2013electron} and thermoelectrics \cite{zhong2012effects}. 
However, the planar growth of III-V-Bi compounds such as GaAs$_{1-\text{x}}$Bi$_{\text{x}}$ is complicated by the requirements of low growth temperatures and group-III-rich growth conditions \cite{noreika1982indium,ma1989organometallic,tixier2003molecular,lewis2012growth,zhong2012effects}. Such conditions often result in the formation of metallic droplets on the surface, which are highly detrimental for device applications \cite{young2007bismuth,vardar2013mechanisms,masnadi2014bandgap}.

Nanowires (NWs) are ideally suited as nanoscale substrates for the growth of highly lattice-mismatched and metastable heterostructures. These structures have made it possible to form coherent heterostructures with materials which would be extremely defective in planar form \cite{lauhon2002epitaxial,bjork2002one,liang2005critical,glas2006critical,kavanagh2010misfit}. Regarding III-V-Bi NWs, the growth of lattice-mismatched GaAs/GaAs$_{1-\text{x}}$Bi$_{\text{x}}$ core–shell NWs has only recently been attempted \cite{Ishikawa2015metamorphic}. Bi has also recently been used as a surfactant during NW growth to alter the crystal structure of Au-catalyzed GaAs NWs from wurtzite to zincblende \cite{lu2014bismuth}, and to induce the self-assembly of InAs quantum dots on GaAs NW sidewalls \cite{lewis2017self}. Additionally, the Bi-catalyzed vapor-liquid-solid (VLS) growth of GaAs nanodiscs \cite{dejarld2014droplet} and of in-plane Ga(As,Bi) NWs \cite{essouda2015bismuth} has been observed. However, the incorporation of Bi in axial NW heterostructures—as well as in VLS growth of standing III-V NWs—has not been reported.

In this work, we exploit the Ga-rich environment at the tip of self-assisted VLS GaAs NWs, to realize GaAs$_{1-\text{x}}$Bi$_{\text{x}}$ axial NW segments with Bi contents in excess of 10$\%$ on Si substrates. Exposing GaAs NWs to Bi results in an accumulation of Bi only within the Ga droplets. Subsequently exposing these Ga-Bi droplets to an As$_2$ flux precipitates axial GaAs$_{1-\text{x}}$Bi$_{\text{x}}$ NW segments, with excess Bi remaining on top of the segment as metal droplets. Varying the temperature during precipitation between 270$\,^{\circ}$C and 380$\,^{\circ}$C, Bi contents up to (10$\pm$2)$\%$ are achieved in the segment. Microstructural and elemental characterization reveals that the GaAs$_{1-\text{x}}$Bi$_{\text{x}}$ segments with highest Bi concentration are single crystalline zincblende. These results illustrate how VLS NW synthesis can be exploited to realize heterostructures with metastable compounds.

\section{Experimental}

\subsection{Sample growth}
Samples were grown by molecular beam epitaxy on p-type Si(111) wafers covered by native oxide. Standard effusion cells provided Ga and Bi fluxes, and valved crackers supplied As$_2$. Substrate temperatures were measured by a pyrometer calibrated to the oxide desorption temperature of GaAs(100). Prior to growth, the wafers were ramped to about 645$\,^{\circ}$C and annealed for 10$\,$min. Ga droplets were then formed on the substrate by exposing the substrate to Ga for 30$\,$s at 1.32$\,$µmh$^{-1}$ GaAs equivalent growth rate, followed by a 220$\,$s flux interruption \cite{kupers2017ga}. Vertical GaAs NW segments with diameter, length and density on the order of 30$\,$nm, 300--700$\,$nm and 0.1--2.5$\,$µm$^{-2}$, respectively, were obtained through simultaneous exposure of As$_2$ at 3.8$\,$µmh$^{-1}$ and Ga at 0.3$\,$µmh$^{-1}$ GaAs equivalent growth rate for 3$\,$min. Subsequently, the As$_2$ flux was decreased to 0.6$\,$µmh$^{-1}$ GaAs equivalent growth rate in order to enlarge the diameter of the NWs \cite{kupers2018diameter}, and the growth was continued for 10$\,$min while maintaining the substrate at 645$\,^{\circ}$C. The resulting GaAs NWs had a total length of 0.3--1.8$\,$µm and a top diameter of 50--110$\,$nm. After the growth of GaAs NWs, the fluxes were interrupted and the substrate ramped to 420$\,^{\circ}$C, and the sample was subsequently exposed to a Bi flux corresponding to a 0.2$\,$µmh$^{-1}$ growth rate for 3\,min to enrich the Ga droplets with Bi. The substrate was then ramped to 270--380$\,^{\circ}$C without the exposure to any fluxes. Axial GaAs$_{1-\text{x}}$Bi$_{\text{x}}$ NW segments were formed with a 30$\,$min exposure of the Ga-Bi droplets to As$_2$ at 0.5$\,$µmh$^{-1}$ GaAs equivalent growth rate. The growth of the samples depicted in figures~\ref{figure1}a,b was stopped after the formation of GaAs NWs, and Ga-Bi droplets, respectively. Samples were rotated at 10$\,$rpm during growth. 

\subsection{Transmission electron microscopy}
Transmission electron microscopy - energy dispersive X-ray spectroscopy (TEM-EDS) studies and morphological analysis were carried out on dispersed NWs. The samples were prepared by mechanically scratching the NWs from their substrates and subsequently transferring them to lacey carbon films with 300 mesh copper grids. TEM and scanning transmission electron microscopy (STEM) imaging, high-resolution TEM imaging, correlated EDS and mapping measurements were performed with a JEOL 2100F field emission microscope operated at 200$\,$kV. The microscope is equipped with a Gatan Ultra Scan 4000 CCD camera for image recording. High-resolution images were obtained by aligning the NWs along the $[1\overline{1}0]$ zone axis orientation, which was controlled by an electron diffraction pattern.

\textbf{g}$_{002}$ dark-field (DF) TEM investigations were carried out on cross-sectional TEM specimens, using a Jeol JEM 3010 microscope operating at 300$\,$kV equipped with a GATAN slow-scan CCD camera. The NWs were prepared in the two orthogonal $[1\overline{1}0]$ and $[11\overline{2}]$ projections using conventional techniques, where the NWs are glued for mechanical stabilization and thinned by mechanical grinding, dimpling, and Ar-ion milling.

\section{Results and discussion}
\begin{figure}[h!]
	\centering
	\includegraphics{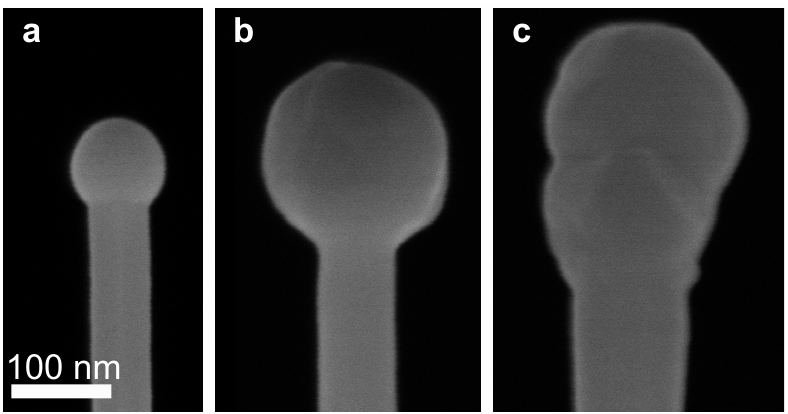}
	\caption[] {Scanning electron micrographs of single NWs after different stages of growth. (a) GaAs NW with Ga droplet. (b) GaAs NW with Bi-Ga droplet with Bi/Ga volume ratio of about 5 after Bi exposure. (c) After exposing the Ga-Bi droplet to As$_2$. The scale bar in (a) is valid for all panels. } 
	\label{figure1}
\end{figure}

Figure~\ref{figure1} depicts the end portion of individual NWs after different stages of growth.
First, GaAs NWs were grown by the Ga-assisted VLS method in molecular beam epitaxy \cite{Colombo2008,jabeen2008self} (figure~\ref{figure1}a). Next, the NWs were exposed to a Bi flux corresponding to a 0.2$\,$µmh$^{-1}$ growth rate of metallic Bi for 3$\,$min at a temperature of about 420$\,^{\circ}$C (figure~\ref{figure1}b). At these conditions, Bi has been shown to behave as a surfactant for growth on GaAs$\lbrace110\rbrace$ NW sidewalls and planar substrates \cite{lewis2017self,lewis2017quantum}. For the present case, assuming a constant ratio between a GaAs NW diameter and the diameter of its Ga droplet, we estimate that based on the analysis of 14 NWs by scanning electron microscopy (SEM), the Bi exposure increases the averaged diameter of the droplets by a factor of about 1.7, indicating an enrichment of the Ga droplet with Bi. This increase corresponds to a Bi/Ga volume ratio in the droplet of about 5. No evidence for Bi accumulation was observed by SEM on the NW sidewalls or substrate, indicating that Bi desorbed from these surfaces [see Supporting Information (SI)]. After enriching the Ga droplets with Bi, the NWs were exposed for 30$\,$min to an As$_2$ flux of 0.5$\,$µmh$^{-1}$ GaAs equivalent planar growth rate at substrate temperatures ranging from 270$\,^{\circ}$C to 380$\,^{\circ}$C. The resulting NW structure obtained for a substrate temperature of 300$\,^{\circ}$C is shown in figure~\ref{figure1}c. The As$_2$ exposure results in the formation of a precipitated segment below the droplet (compare figure~\ref{figure1}b and~\ref{figure1}c). 

According to the bulk Ga-Bi phase diagram, Ga and Bi form a homogeneous mixture for all Bi concentrations above the Bi melting point (271.4$\,^{\circ}$C) \cite{Predel1960,elayech2015}. Thus, the Ga-Bi droplet should be a homogeneous liquid mixture for all As$_2$ exposure temperatures, with the possible exception of the lowest temperature of 270$\,^{\circ}$C. The addition of As$_2$ is expected to precipitate a GaAs-rich solid \cite{leonhardt1974loslichkeiten,evgenev1984phase}. We note that GaAs/InAs axial NW heterostructures have previously been obtained by a related approach, where In droplets formed atop of GaAs NWs (where the Ga droplets had previously been converted to GaAs by As exposure) were subsequently converted into InAs axial segments by As exposure \cite{somaschini2014axial,Scarpellini2015}.

\begin{figure}[h!]
	\centering
	\includegraphics{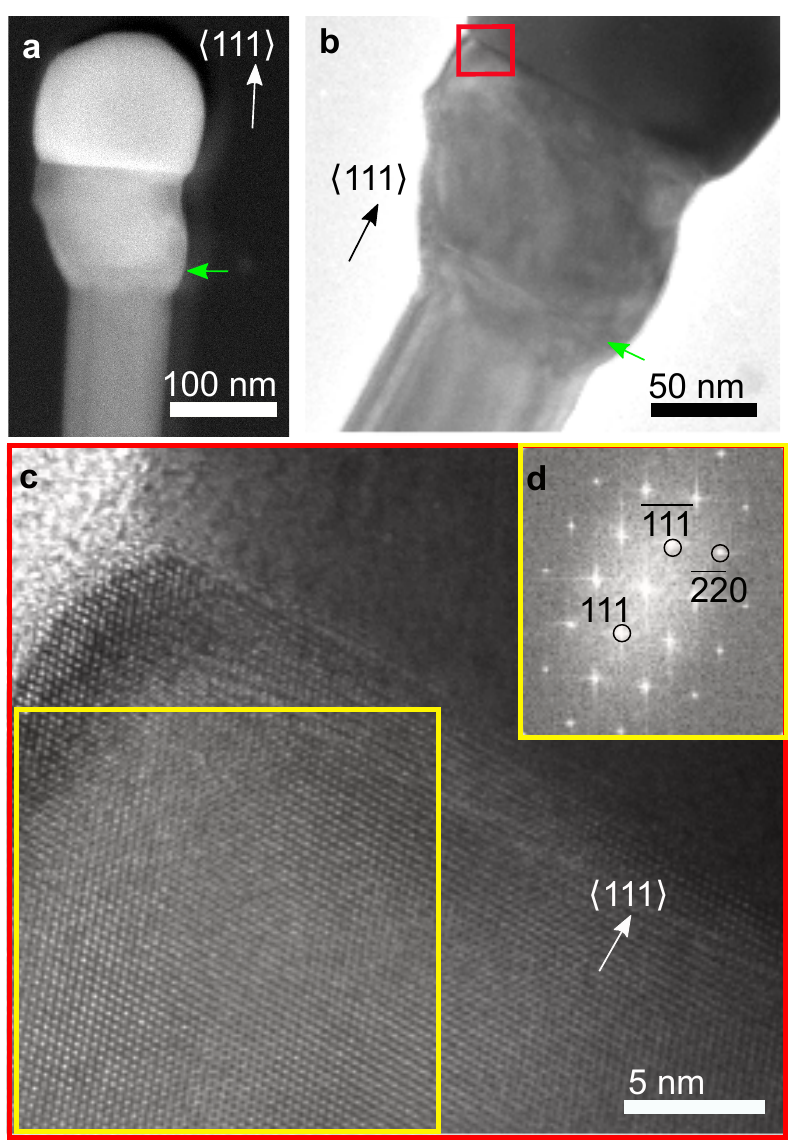}
	\caption[] {TEM micrographs of a single axial GaAs/GaAs$_{1-\text{x}}$Bi$_{\text{x}}$ NW heterostructure grown by exposing the Ga-Bi droplet to As$_2$ at 300$\,^{\circ}$C. (a) High-angle annular dark-field image. (b) Bright-field image. (c) High-resolution TEM microfield of the red-highlighted droplet-NW interface region in (b). (d) Fast Fourier transform of the yellow heightened region corresponding to the $[1\overline{1}0]$ zone axis in (c), indicating single crystal zincblende structure.  The NW was aligned close to the $[1\overline{1}0]$ zone axis, and its growth direction was along $\braket{111}$, most likely along $[\overline{1}\overline{1}\overline{1}]$ \cite{i2011gold,lin2017efficient}. Green arrows indicate the apparent GaAs/GaAs$_{1-\text{x}}$Bi$_{\text{x}}$ interface. }
	\label{figure2}
\end{figure}

The microstructure of NWs grown by the above approach was investigated by TEM. Figure~\ref{figure2} shows TEM micrographs of a single GaAs/GaAs$_{1-\text{x}}$Bi$_{\text{x}}$ NW hererostructure grown by exposing the Ga-Bi droplet to As$_2$ at 300$\,^{\circ}$C [see SI for micrographs on additional NWs]. 
The high angle annular dark-field (HAADF) STEM image of the NW (figure~\ref{figure2}a) illustrates three regions of distinct contrast: a lower region corresponding to the GaAs NW stem, a middle region which precipitated from the Ga-Bi droplet during the As$_2$ exposure step, and a remaining top droplet. 

Analysis of 15 NWs from this sample yields average widths of (95$\pm$3)$\,$nm for the lower GaAs segment and (147$\pm$5)$\,$nm for the precipitated middle segment. As NW widths are correlated with the droplet diameter \cite{cui2001diameter}, the wider diameter is consistent with this segment having precipitated from the Ga-Bi droplet, which was larger than the initial Ga droplet. The precipitated segment shows a brighter contrast than the lower GaAs in figure~\ref{figure2}a, due to the differences in thickness and/or change in composition due to the presence of Bi. As the HAADF intensity is proportional to the atomic number to the power of about 1.7, regions containing Bi (atomic number 83) are expected to appear brighter than those containing only Ga and As (atomic numbers 31 and 33, respectively). Correspondingly, the droplet atop the NW shows the brightest contrast, indicating that it is rich in Bi.

A bright-field TEM micrograph of the NW shown in figure~\ref{figure2}a is presented in figure~\ref{figure2}b. An abrupt change in contrast in the axial direction is observed near the bottom of the precipitated segment in both figure~\ref{figure2}a and~\ref{figure2}b (indicated by green arrows in both images), possibly indicating an interface between GaAs and GaAs$_{1-\text{x}}$Bi$_{\text{x}}$. Figure~\ref{figure2}c displays a close-up high-resolution TEM micrograph of the interface between the droplet and the precipitated segment (red-highlighted area in figure~\ref{figure2}b), revealing an abrupt interface between the two regions. While the precipitated segment exhibits a single crystal zincblende crystal structure (see figure~\ref{figure2}d showing a fast Fourier transform of the area indicated in yellow), the droplet in figure~\ref{figure2}c shows no clear crystal structure. \\

\begin{figure}[h!]
\centering
\includegraphics{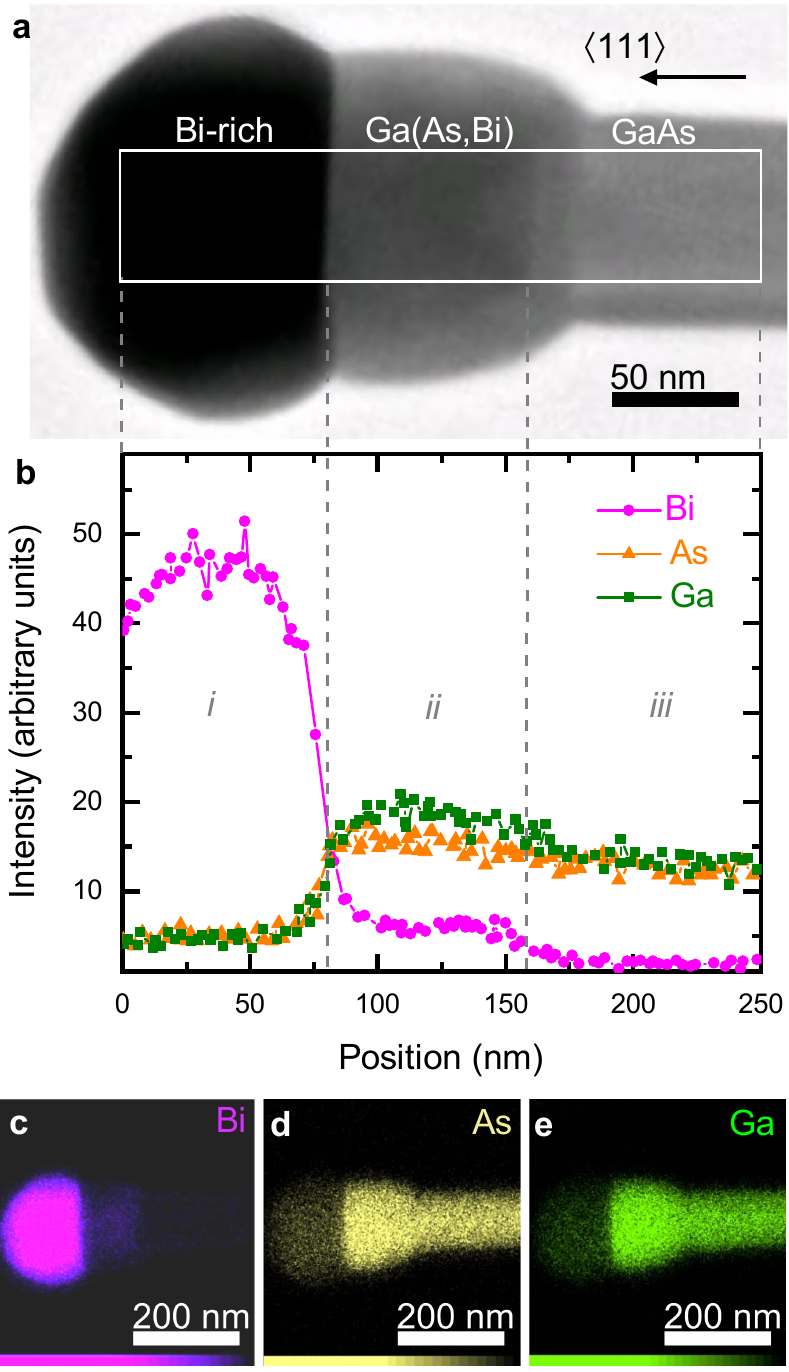}
\caption[] {TEM-EDS characterization of a single axial GaAs/GaAs$_{1-\text{x}}$Bi$_{\text{x}}$ NW heterostructure with GaAs$_{1-\text{x}}$Bi$_{\text{x}}$ segment grown at 300$^{\circ}\,\text{C}$. (a) Bright-field micrograph (growth direction indicated by arrow). (b) EDS line profile measured over the highlighted area in (a). The plot is separated into three regions labeled \textit{i}, \textit{ii} and \textit{iii}, which correspond to the Bi-rich droplet, the GaAs$_{1-\text{x}}$Bi$_{\text{x}}$ and the GaAs segments, respectively. (c--e) EDS maps for Bi, As and Ga.} 
\label{figure3}
\end{figure}

The local elemental composition of individual NW heterostructures was investigated by TEM-EDS. Figure~\ref{figure3} presents the TEM-EDS charactization of the composition of a single NW with a GaAs$_{1-\text{x}}$Bi$_{\text{x}}$ segment grown at 300$^{\circ}\,\text{C}$ [see SI for TEM-EDS data of additional NWs]. A bright field TEM micrograph of the investigated NW is shown in figure~\ref{figure3}a. Figure~\ref{figure3}b presents horizontal EDS line profiles of Bi, As and Ga, carried out over the area indicated by the white rectangle in~\ref{figure3}a. Gray dotted lines separate three distinct regions of composition:  Bi-rich droplet (\textit{i}), GaAs$_{1-\text{x}}$Bi$_{\text{x}}$ segment (\textit{ii}) and GaAs NW (\textit{iii}), which correspond well with the features in the bright-field micrograph. Region \textit{i} is dominated by the presence of Bi, but also consists of a non-zero amount of As and Ga. Region \textit{ii} contains significant concentrations of Bi, As and Ga, thus demonstrating the successful incorporation of Bi in an axial GaAs$_{1-\text{x}}$Bi$_{\text{x}}$ NW segment. The Bi signal is remarkably constant in the interval 100--150$\,$nm. Region \textit{iii} is essentially Bi-free, as expected for the GaAs segment of the NW. We note that the abrupt contrast change observed in figure~\ref{figure3}a corresponds to an abrupt increase in the Bi content. 
Such an abrupt contrast is also observed for the NW shown in figures~\ref{figure2}a,b (indicated by green arrows). Figure~\ref{figure3}c--e show the corresponding two-dimensional maps of Bi, As, and Ga for the NW heterostructure. The EDS results indicate that exposing the Ga-Bi droplet to As$_2$ results in the precipitation of a GaAs$_{1-\text{x}}$Bi$_{\text{x}}$ segment. This process depletes the droplet of most of the Ga. The remaining Bi that does not incorporate into the precipitated segment remains as a Bi-rich droplet.

Since the EDS intensity depends on the composition, element specific scattering probability and thickness, the Bi, As and Ga intensities in~\ref{figure3}b are not directly comparable. However, we note that in region \textit{iii}, which is expected to contain stoichiometric GaAs, the Ga and As signals are nearly equal. By contrast, in region \textit{ii} the Ga signal is considerably larger than the As signal. Since Bi is expected to occupy the group V (As) sites in the GaAs lattice, the relative drop in the As signal in region \textit{ii} is attributed to Bi incorporation on As sites. Under this assumption, the As content in the GaAs$_{1-\text{x}}$Bi$_{\text{x}}$ segment (\textit{ii}), 1-x, is given by

 \begin{equation}
 [\mtext{As}]_{\mtext{\textit{ii}}} = \left(\mtext{I}_{\text{As}}/ \mtext{I}_{\text{Ga}}\right)_{\text{\textit{ii}}} \cdot \left(\mtext{I}_{\text{Ga}}/ \mtext{I}_{\text{As}} \right)_{\text{\textit{iii}}},
\label{eq}
 \end{equation}

where $\mtext{I}_{\text{As}}$, $\mtext{I}_{\text{Ga}}$ are the As and Ga intensities, respectively, in the indicated region (\textit{ii}, \textit{iii}). Thus, the As/Ga intensity ratio in region \textit{ii} is normalized to that of region \textit{iii}, which is assumed to contain stoichiometric GaAs. The shown data and a repeated scan of the same NW yield an averaged Bi concentration, x, of (10$\pm$2)$\%$. For comparison, we note that the Bi signal in the GaAs$_{1-\text{x}}$Bi$_{\text{x}}$ segment is on the order of 10$\%$ of that in the droplet. Since the droplet is composed of predominantly Bi, this further confirms a large incorporation of Bi in the GaAs$_{1-\text{x}}$Bi$_{\text{x}}$ segment on the order of 10$\%$. 

\begin{figure}[h!]
\centering
\includegraphics{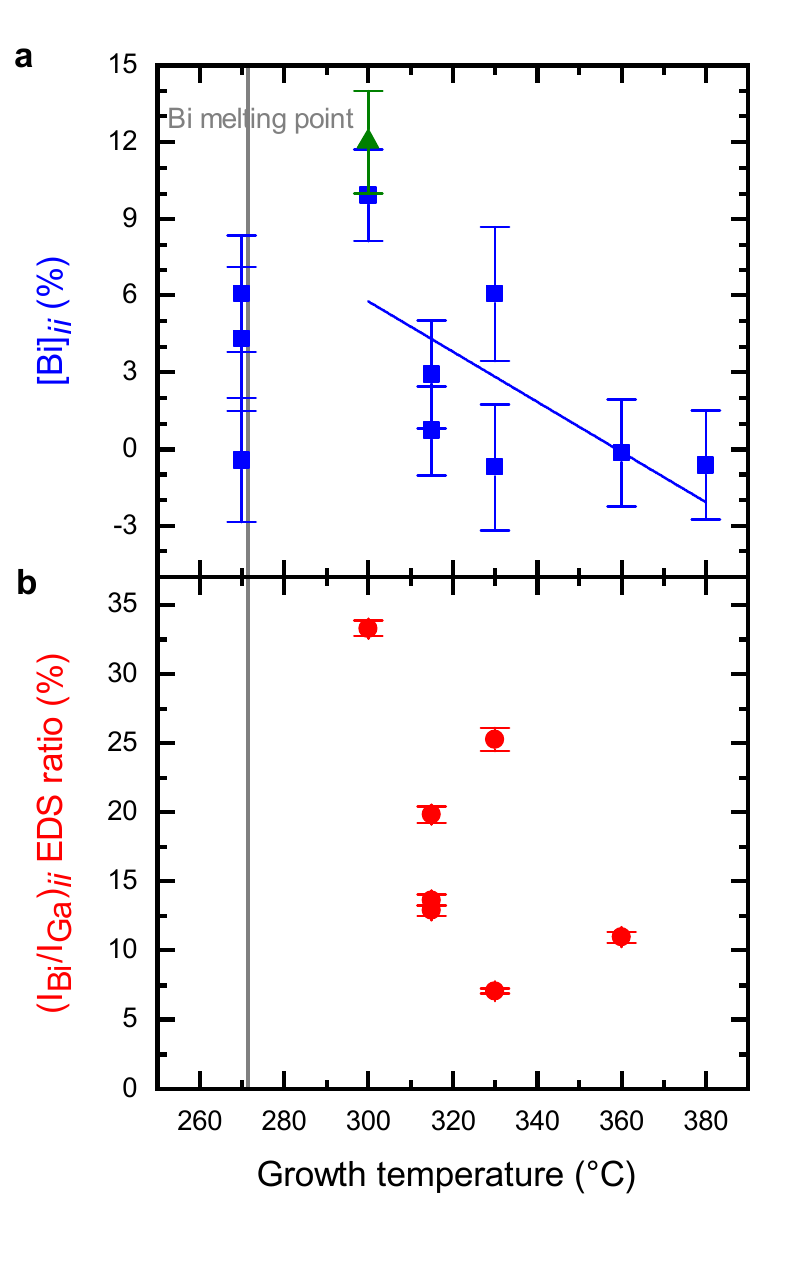}
\caption[] {(a) GaAs$_{1-\text{x}}$Bi$_{\text{x}}$ segment Bi content as a function of the growth temperature during the As$_2$ exposure step, determined by the TEM-EDS method discussed in the text (blue data). A linear fit to the data for growth temperatures 300--380$^{\circ}\,\text{C}$ is included to illustrate the trend of the data. Negative values should not be associated with physical meaning. The green data point was obtained with independent \textbf{g}$_{002}$ dark-field TEM measurements. (b) Bi/Ga EDS intensity ratio of GaAs$_{1-\text{x}}$Bi$_{\text{x}}$ segments. Data points correspond to measurements on individual NWs. The gray line indicates the bulk melting point of Bi.}
\label{figure4}
\end{figure}

The above procedure was carried out on other samples, where the GaAs$_{1-\text{x}}$Bi$_{\text{x}}$ segment was precipitated at different substrate temperatures. Figure~\ref{figure4}a shows the Bi content $[\mtext{Bi}]_{\text{\textit{ii}}}$ of GaAs$_{1-\text{x}}$Bi$_{\text{x}}$ segments as a function of growth temperature. The data points are obtained from measurements on single NWs [TEM-EDS data is provided in the SI]. We note that due to measurement uncertainties, this procedure may result in negative values for $[\mtext{Bi}]_{\text{\textit{ii}}}$, which should not be associated with physical meaning.

The Bi concentration strongly decreases with increasing growth temperature for temperatures above 300$\,^{\circ}$C, in agreement with planar GaAs$_{1-\text{x}}$Bi$_{\text{x}}$ growth studies \cite{lewis2012growth,ptak2012kinetically}. The highest Bi content of (10$\pm$2)$\%$ was obtained at 300$\,^{\circ}$C. At this substrate temperature, Bi concentrations around 4$\%$ \cite{ptak2012kinetically} and 8$\%$ \cite{lewis2012growth} were achieved in planar layers. Thus, it is possible to incorporate Bi into axial GaAs$_{1-\text{x}}$Bi$_{\text{x}}$ NW segments in very high concentrations. Interestingly, we observe a scatter in the Bi content of single NWs (even from NWs of the same sample), possibly due to local variations in the environment around the individual Ga-Bi droplets during the precipitation of the GaAs$_{1-\text{x}}$Bi$_{\text{x}}$ segments. We speculate that this fluctuation could be minimized by employing selective area growth to realize regular arrays of GaAs NWs \cite{Plissard2011,kupers2017surface}. For segments grown at 270$\,^{\circ}$C, TEM micrographs [see SI] reveal inhomogeneous morphologies as well as the presence of clusters. We note that the melting point of Bi is 271.4$\,^{\circ}$C. It is probable that these segments precipitated from Bi-Ga droplets which were partially solid, which presumably would strongly influence the segment growth.

To corroborate the Bi concentrations determined from the As and Ga EDS signals, the average Bi/Ga EDS intensity ratio $\left(\mtext{I}_{\text{Bi}}/\mtext{I}_{\text{Ga}}\right)_{\text{\textit{ii}}}$ of the GaAs$_{1-\text{x}}$Bi$_{\text{x}}$ segments is plotted as a function of growth temperature in figure~\ref{figure4}b. We note that these values are not the Bi concentration in the GaAs$_{1-\text{x}}$Bi$_{\text{x}}$ segment, as EDS intensities scattered by different elements are not directly comparable. However, we expect that the obtained values only differ from the real Bi content by a constant scale factor. We note that segments with higher Bi concentrations in figure~\ref{figure4}a, correspond to higher $\left(\mtext{I}_{\text{Bi}}/\mtext{I}_{\text{Ga}}\right)_{\textit{ii}}$ ratios in~\ref{figure4}b, and the $\left(\mtext{I}_{\text{Bi}}/\mtext{I}_{\text{Ga}}\right)_{\textit{ii}}$ ratio also strongly decreases with increasing growth temperature. The strong agreement between $[\mtext{Bi}]_{\textit{ii}}$ and $\left(\mtext{I}_{\text{Bi}}/\mtext{I}_{\text{Ga}}\right)_{\textit{ii}}$ indicates that the large NW-to-NW Bi content fluctuations are real and not the result of measurement inaccuracies. In order to calculate the $\left(\mtext{I}_{\text{Bi}}/\mtext{I}_{\text{Ga}}\right)_{\textit{ii}}$ ratio from TEM-EDS data, it is necessary that the Bi EDS signal is unequivocally attributable to Bi integrated in the GaAs$_{1-\text{x}}$Bi$_{\text{x}}$ segment, and not to Bi-rich accumulations on the side of the GaAs$_{1-\text{x}}$Bi$_{\text{x}}$ segment. For this reason, the $\left(\mtext{I}_{\text{Bi}}/\mtext{I}_{\text{Ga}}\right)_{\text{\textit{ii}}}$ ratio is not given for some of the NWs [see SI]. 

Additionally, independent chemically sensitive \textbf{g}$_{002}$ DFTEM measurements \cite{wu2015} adapted to the NW geometry allow a rough estimation of the Bi content [for details, see the SI]. The measurements were performed on the NWs with GaAs$_{1-\text{x}}$Bi$_{\text{x}}$ segments grown at 300$\,^{\circ}$C. The analysis yields an average Bi content of [Bi] = (12$\pm$2)$\%$ (see green data in figure~\ref{figure4}a), which is in good agreement with the value estimated using EDS.  

The growth of GaAs$_{1-\text{x}}$Bi$_{\text{x}}$ planar films with Bi-contents in excess of 10$\%$ is challenging owing to the weak Ga-Bi reactivity and the requirement of Ga-rich growth conditions \cite{lewis2012growth}. The bandgap of GaAs$_{1-\text{x}}$Bi$_{\text{x}}$ films with about 10$\%$ Bi has been shown to be about 0.8~eV \cite{lu2009composition,masnadi2014bandgap}. Additionally, it has been shown that GaAs$_{1-\text{x}}$Bi$_{\text{x}}$ planar layers with Bi contents of up to 17.8$\%$ have a direct bandgap \cite{masnadi2014bandgap}. Our axial GaAs$_{1-\text{x}}$Bi$_{\text{x}}$/GaAs NW heterostructures are thus of interest for optoelectronic devices, potentially covering a huge bandgap range of 0.8--1.424$\,$eV, which includes the telecommunications windows.

\section{Summary and conclusions}
In summary, axial GaAs$_{1-\text{x}}$Bi$_{\text{x}}$/GaAs NW heterostructures with high Bi concentrations have been realized by exploiting the Ga-rich environment atop self-assisted GaAs NWs. This growth approach overcomes the major challenges which have faced the planar growth of the exciting family of metastble III-V-Bi alloys. Characterization of single GaAs$_{1-\text{x}}$Bi$_{\text{x}}$/GaAs NW heterostructures demonstrates the formation of single crystalline, zincblende GaAs$_{1-\text{x}}$Bi$_{\text{x}}$ with Bi concentrations up to about (10$\pm$2)$\%$. This work demonstrates how the local environment present during NW growth can be exploited to synthesize metastable alloys at conditions which cannot be reached by more conventional growth approaches. 

\ack 

The authors acknowledge M. Höricke and C. Stemmler for MBE maintenance, U. Jahn, M. Ramsteiner, C. Sinito and J. Lähnemann for experimental support and discussions, O. Brandt and A. Trampert for invaluable discussions, and K. Biermann for a critical reading of the manuscript. R.\,B. Lewis acknowledges funding from the Alexander von Humboldt foundation.


\bibliographystyle{iopart-num}
\bibliography{Oliva-GaAsBi2019}

\providecommand{\newblock}{}
\begin{thebibliography}{10}
\expandafter\ifx\csname url\endcsname\relax
  \def\url#1{{\tt #1}}\fi
\expandafter\ifx\csname urlprefix\endcsname\relax\def\urlprefix{URL }\fi
\providecommand{\eprint}[2][]{\url{#2}}

\bibitem{francoeur2003band}
Francoeur S, Seong M~J, Mascarenhas A, Tixier S, Adamcyk M and Tiedje T 2003
  {\em Appl. Phys. Lett.\/} {\bf 82} 3874--3876

\bibitem{Marko2012}
Marko I, Batool Z, Hild K, Jin S, Hossain N, Hosea T, Petropoulos J, Zhong Y,
  Dongmo P, Zide J {\em et~al.\/} 2012 {\em Appl. Phys. Lett.\/} {\bf 101}
  221108

\bibitem{zhong2012effects}
Zhong Y, Dongmo P, Petropoulos J and Zide J 2012 {\em Appl. Phys. Lett.\/} {\bf
  100} 112110

\bibitem{rajpalke2014high}
Rajpalke M~K, Linhart W, Birkett M, Yu K, Alaria J, Kopaczek J, Kudrawiec R,
  Jones T, Ashwin M and Veal T 2014 {\em J. Appl. Phys.\/} {\bf 116} 043511

\bibitem{sandall2014demonstration}
Sandall I, Bastiman F, White B, Richards R, Mendes D, David J and Tan C 2014
  {\em Appl. Phys. Lett.\/} {\bf 104} 171109

\bibitem{masnadi2014bandgap}
Masnadi-Shirazi M, Lewis R~B, Bahrami-Yekta V, Tiedje T, Chicoine M and Servati
  P 2014 {\em J. Appl. Phys.\/} {\bf 116} 223506

\bibitem{fluegel2006giant}
Fluegel B, Francoeur S, Mascarenhas A, Tixier S, Young E and Tiedje T 2006 {\em
  Phys. Rev. Lett.\/} {\bf 97} 067205

\bibitem{mazzucato2013electron}
Mazzucato S, Zhang T, Carr{\`e}re H, Lagarde D, Boonpeng P, Arnoult A, Lacoste
  G, Balocchi A, Amand T, Fontaine C and Marie X 2013 {\em Appl. Phys. Lett.\/}
  {\bf 102} 252107

\bibitem{noreika1982indium}
Noreika A, Takei W, Francombe M and Wood C 1982 {\em J. Appl. Phys.\/} {\bf 53}
  4932--4937

\bibitem{ma1989organometallic}
Ma K, Fang Z, Jaw D, Cohen R, Stringfellow G, Kosar W and Brown D 1989 {\em
  Appl. Phys. Lett.\/} {\bf 55} 2420--2422

\bibitem{tixier2003molecular}
Tixier S, Adamcyk M, Tiedje T, Francoeur S, Mascarenhas A, Wei P and
  Schiettekatte F 2003 {\em Appl. Phys. Lett.\/} {\bf 82} 2245--2247

\bibitem{lewis2012growth}
Lewis R, Masnadi-Shirazi M and Tiedje T 2012 {\em Appl. Phys. Lett.\/} {\bf
  101} 082112

\bibitem{young2007bismuth}
Young E, Whitwick M, Tiedje T and Beaton D 2007 {\em Phys. Status Solidi C\/}
  {\bf 4} 1707--1710

\bibitem{vardar2013mechanisms}
Vardar G, Paleg S, Warren M, Kang M, Jeon S and Goldman R 2013 {\em Appl. Phys.
  Lett.\/} {\bf 102} 042106

\bibitem{lauhon2002epitaxial}
Lauhon L~J, Gudiksen M~S, Wang D and Lieber C~M 2002 {\em Nature\/} {\bf 420}
  57

\bibitem{bjork2002one}
Bj{\"o}rk M, Ohlsson B, Sass T, Persson A, Thelander C, Magnusson M, Deppert K,
  Wallenberg L and Samuelson L 2002 {\em Nano Lett.\/} {\bf 2} 87--89

\bibitem{liang2005critical}
Liang Y, Nix W~D, Griffin P~B and Plummer J~D 2005 {\em J. Appl. Phys.\/} {\bf
  97} 043519

\bibitem{glas2006critical}
Glas F 2006 {\em Phys. Rev. B: Condens. Matter Mater. Phys.\/} {\bf 74} 121302

\bibitem{kavanagh2010misfit}
Kavanagh K~L 2010 {\em Semicond. Sci. Technol.\/} {\bf 25} 024006

\bibitem{Ishikawa2015metamorphic}
Ishikawa F, Akamatsu Y, Watanabe K, Uesugi F, Asahina S, Jahn U and Shimomura S
  2015 {\em Nano Lett.\/} {\bf 15} 7265--7272

\bibitem{lu2014bismuth}
Lu Z, Zhang Z, Chen P, Shi S, Yao L, Zhou C, Zhou X, Zou J and Lu W 2014 {\em
  Appl. Phys. Lett.\/} {\bf 105} 162102

\bibitem{lewis2017self}
Lewis R~B, Corfdir P, Herranz J, K{\"u}pers H, Jahn U, Brandt O and Geelhaar L
  2017 {\em Nano Lett.\/} {\bf 17} 4255--4260

\bibitem{dejarld2014droplet}
DeJarld M, Nothern D and Millunchick J~M 2014 {\em J. Appl. Phys.\/} {\bf 115}
  114307

\bibitem{essouda2015bismuth}
Essouda Y, Fitouri H, Boussaha R, Elayech N, Rebey A and El~Jani B 2015 {\em
  Mater. Lett.\/} {\bf 152} 298--301

\bibitem{kupers2017ga}
K{\"u}pers H, Bastiman F, Luna E, Somaschini C and Geelhaar L 2017 {\em J.
  Cryst. Growth\/} {\bf 459} 43--49

\bibitem{kupers2018diameter}
K{\"u}pers H, Lewis R~B, Tahraoui A, Matalla M, Kr{\"u}ger O, Bastiman F,
  Riechert H and Geelhaar L 2018 {\em Nano Res.\/} {\bf 11} 2885--2893

\bibitem{Colombo2008}
Colombo C, Spirkoska D, Frimmer M, Abstreiter G and Morral A~F~i 2008 {\em
  Phys. Rev. B: Condens. Matter Mater. Phys.\/} {\bf 77} 155326

\bibitem{jabeen2008self}
Jabeen F, Grillo V, Rubini S and Martelli F 2008 {\em Nanotechnology\/} {\bf
  19} 275711

\bibitem{lewis2017quantum}
Lewis R~B, Corfdir P, Li H, Herranz J, Pf{\"u}ller C, Brandt O and Geelhaar L
  2017 {\em Phys. Rev. Lett.\/} {\bf 119} 086101

\bibitem{Predel1960}
Predel B 1960 {\em Zeitschrift für Physikalische Chemie\/} {\bf 24} 206--216

\bibitem{elayech2015}
Elayech N, Fitouri H, Essouda Y, Rebey A and El~Jani B 2015 {\em Phys. Status
  Solidi C\/} {\bf 12} 138--141

\bibitem{leonhardt1974loslichkeiten}
Leonhardt A and K{\"u}hn G 1974 {\em Krist. Tech.\/} {\bf 9} 77--85

\bibitem{evgenev1984phase}
Evgenev S and Ganina N 1984 {\em Inorg. Mater.\/} {\bf 20} 479--481

\bibitem{somaschini2014axial}
Somaschini C, Biermanns A, Bietti S, Bussone G, Trampert A, Sanguinetti S,
  Riechert H, Pietsch U and Geelhaar L 2014 {\em Nanotechnology\/} {\bf 25}
  485602

\bibitem{Scarpellini2015}
Scarpellini D, Somaschini C, Fedorov A, Bietti S, Frigeri C, Grillo V, Esposito
  L, Salvalaglio M, Marzegalli A, Montalenti F, Bonera E, Medaglia P~G and
  Sanguinetti S 2015 {\em Nano Lett.\/} {\bf 15} 3677--3683

\bibitem{i2011gold}
i~Morral A~F 2011 {\em IEEE J. Sel. Top. Quantum Electron.\/} {\bf 17} 819--828

\bibitem{lin2017efficient}
Lin W~H, Jahn U, K{\"u}pers H, Luna E, Lewis R~B, Geelhaar L and Brandt O 2017
  {\em Nanotechnology\/} {\bf 28} 415703

\bibitem{cui2001diameter}
Cui Y, Lauhon L~J, Gudiksen M~S, Wang J and Lieber C~M 2001 {\em Appl. Phys.
  Lett.\/} {\bf 78} 2214--2216

\bibitem{ptak2012kinetically}
Ptak A, France R, Beaton D, Alberi K, Simon J, Mascarenhas A and Jiang C~S 2012
  {\em J. Cryst. Growth\/} {\bf 338} 107--110

\bibitem{Plissard2011}
Plissard S~R, Larrieu G, Wallart X and Caroff P 2011 {\em Nanotechnology\/}
  {\bf 22} 275602

\bibitem{kupers2017surface}
K{\"u}pers H, Tahraoui A, Lewis R~B, Rauwerdink S, Matalla M, Kr{\"u}ger O,
  Bastiman F, Riechert H and Geelhaar L 2017 {\em Semicond. Sci. Technol.\/}
  {\bf 32} 115003

\bibitem{wu2015}
Wu M, Hanke M, Luna E, Puustinen J, Guina M and Trampert A 2015 {\em
  Nanotechnology\/} {\bf 26} 425701

\bibitem{lu2009composition}
Lu X, Beaton D, Lewis R~B, Tiedje T and Zhang Y 2009 {\em Appl. Phys. Lett.\/}
  {\bf 95} 041903

\end{thebibliography}

\end{document}